\documentstyle[twocolumn,prl,aps,epsfig,amssymb]{revtex}
\def\pslash{\rlap{\hspace{0.02cm}/}{p}}
\begin{document}

\title{Detecting the neutral top-pion at $e^{+}e^{-}$ colliders }
\author{Xuelei Wang , Yueling Yang, and Bingzhong Li}
\address{ College of Physics and Information Engineering, Henan Normal University,
Henan 453002, China}
\date{\today}

\maketitle
\begin{abstract}
\begin{center}
Abstract
\end{center}
 We investigate some processes of  the associated production
of a neutral top-pion $\Pi^{0}_{t}$ with
 a pair of fermions($e^{+}e^{-}\rightarrow  f\overline{f}\Pi^{0}_{t}$) in the context of top-color-assisted
 technicolor(TC2) theory at future $e^{+}e^{-}$ colliders.
 The studies show that the largest cross sections
 of the processes  $e^{+}e^{-}\rightarrow  f'\overline{f'}\Pi^{0}_{t}(f'=u,d,c,s,\mu,\tau)$
 could only reach the
 level of  0.01fb, we can hardly detect a neutral top-pion through
 these processes.  For the  processes  $e^{+}e^{-}\rightarrow
 e^{+}e^{-}\Pi^{0}_{t}$, $e^{+}e^{-}\rightarrow
 t\overline{t}\Pi^{0}_{t}$ and $e^{+}e^{-}\rightarrow
 b\overline{b}\Pi^{0}_{t}$,
 the cross sections of these processes are  at the level of
 a few fb for the favorable parameters and a few tens, even
 hundreds, of
  neutral top-pion events can be produced at future $e^{+}e^{-}$ colliders each year through these processes.
  With the clean background of the flavor-changing $t\bar{c}$ channel, the top-pion events can possibly be detected
   at the planned  high luminosity $e^{+}e^{-}$ colliders.
   Therefore, such neutral top-pion production processes
   provide a useful way to detect a neutral top-pion
  and test the TC2 model directly.
\end{abstract}

\section{ Introduction}
 Essential elements of fundamental constituents of matter and
 their interactions have been discovered in the past three decades
 by operating $e^{+}e^{-}$ colliders. A coherent  picture of the
 structure of matter has emerged, that is adequately described by
 the standard model(SM), in many of its facets at a level of very
 high accuracy. However, the SM does not provide a comprehensive
 theory of matter. Neither the fundamental parameters, masses and
 couplings, nor the symmetry pattern can be explained. Therefore,
 some new theories beyond the SM(new physics models) have been studied to solve the
 problems of the SM. On the other hand, there exist new particles in
 the new physics models, probing these particles at future high-energy colliders
 can provide a direct way to test these models.

 \subsection{The top-color-assisted technicolor model}
 The top quark is the heaviest particle yet experimentally
 discovered and its mass of 174 GeV \cite{cdf} is close to the
 electroweak symmetry breaking(EWSB) scale. Much theoretical work
 has been carried out in connection to the top quark and EWSB. The
 top-color-assisted technicolor (TC2)\cite{TC2} model, the top seesaw
 model \cite{seesaw} and the flavor universal TC2 model
 \cite{utc2} are three such examples. The TC2 model is a more
 realistic one, which generates the large top quark mass through
 a dynamical $<t\bar{t}>$ condensation and
 provides a possible dynamical mechanism for breaking electroweak
 symmetry. In the TC2 model, the new strong dynamics is assumed to be
 chiral-critically strong but spontaneously broken by TC at the
 scale $\backsim$ 1 TeV and the EWSB is driven mainly by
 TC interaction. The extended technicolor(ETC) interaction gives the contribution
 to all ordinary quark and lepton masses including a very small
 portion of the top quark mass:  $m^{'}_{t}=\varepsilon m_{t}$ $(0.03\leq
\varepsilon \leq 0.1)$\cite{Burdman}. The top-color interaction
also makes a small contribution to the EWSB and gives rise to the
main part of the top quark mass: $(1-\varepsilon)m_{t}$. The TC2
model also predicts the existence  of  a CP-even scalar
$(h^{0}_{t})$ called the top-Higgs   and  three  CP-odd pseudo
Goldstone bosons(PGB's) called top-pions
$(\Pi^{0}_{t}$,$\Pi^{\pm}_{t})$ in a few hundreds GeV region. The
physical particle top-pions can be regarded as a typical feature
of the TC2 model. Thus, the study of the possible signatures of
top-pions and top-pion contribution to some processes at high-
energy colliders can be regarded as a good method to test the TC2
model and further to probe the EWSB mechanism.

 At the energy scale $\Lambda\backsim$ 1 TeV, the new strong dynamics is coupled
preferentially to the third generation. The dynamics of a general
TC2 model involves the following structure\cite{TC2,Burdman}:
\begin{eqnarray*}
SU(3)_{1}\bigotimes SU(3)_{2}\bigotimes U(1)_{Y_{1}}\bigotimes
U(1)_{Y_{2}}& &\bigotimes SU(2)_{L}\\
 \rightarrow SU(3)_{QCD}\bigotimes U(1)_{EM}
\end{eqnarray*}
where $SU(3)_{1}\bigotimes U(1)_{Y_{1}}(SU(3)_{2}\bigotimes
U(1)_{Y_{2}})$ couples preferentially to the third generation(the
first and the second generations). The $U(1)_{Y_{i}}$ is just
strongly rescaled versions of electroweak $U(1)_{Y}$.
$SU(3)_{1}\bigotimes U(1)_{Y_{1}}$ is assumed strong enough to
produce a large top condensate which is responsible for the main
part of the top quark mass. The b-quark mass is an interesting
issue, involving a combination of ETC effects and instanton
effects in $SU(3)_{1}$. The instanton induced b-quark mass can
then be estimated as \cite{Simmons}:
\begin{eqnarray*}
m_{b}^{*}\approx \frac{3km_{t}}{8\pi^{2}}\thicksim 6.6k ~~GeV
\end{eqnarray*}
where we generally expect $k\thicksim 1$ to $10^{-1}$ as in QCD.
In the TC2 model, the top-color gauge bosons include the
color-octet colorons $B^{A}_{\mu}$ and color-singlet extra $U(1)$
gauge boson $Z^{'}$. These gauge bosons have very large masses
which can be up to several TeV. Such large masses will depress the
contribution to the cross sections. So, in our calculation, we can
neglect the contributions of the gauge bosons.

\subsection{Search for the new particles at planned $e^{+}e^{-}$ colliders}
The planned linear $e^{+}e^{-}$ colliders(LC) with energy in the
range from a few hundred GeV up to several TeV are under intense
studies around the world. These studies are being done at the Next
Linear Collider(NLC)(USA)\cite{NLC}, the Japan Linear
Collider(JLC)(Japan)\cite{JLC} and the DESY TeV Energy
Superconducting Linear Accelerator(TESLA)(Europe)\cite{TESLA}. One
task of these high-energy $e^{+}e^{-}$ colliders is to search for
Higgs particle in the SM or some new particles predicted in the
models beyond the SM[such as Higgs bosons
$A^{0},H^{0},h^{0},H^{\pm}$ in the minimal supersymmetric standard
model(MSSM) and PGB's  in the TC2 model]. So, the study of some
new particle production processes can provide a theoretical
instruction to search for these particles experimentally.

 As it is known, top-pions are the typical particles in the TC2 model.
 Some neutral top-pion production processes have been studied in Ref.\cite{wang}.
On the other hand, Ref.\cite{cao} has studied a top-charm
associated production process at LHC to probe the top-pion. The
above studies provide us with some useful information to detect
top-pion events and test the TC2 model. In this paper, we study
the neutral top-pion production
 processes $e^{+}e^{-}\rightarrow f\bar{f}\Pi^{0}_{t}$ in the
 framework of the TC2 model, where $f$ represents $u,d,c,s,t,b$ quarks
 and $e,\mu,\tau$ leptons. Our results show that the cross
 sections
 of the processes $e^{+}e^{-}\rightarrow f'\bar{f'}\Pi^{0}_{t}(f'=u,d,c,s,\mu,\tau)$
 are very
 small. The largest cross section could only reach   an order of
 magnitude $0.01fb$. With such small cross sections, the neutral top-pion can be hardly
 detected via these processes. So, we pay attention to the
 processes $e^{+}e^{-}\rightarrow e^{+}e^{-}\Pi^{0}_{t}$,
 $e^{+}e^{-}\rightarrow t\bar{t}\Pi^{0}_{t}$
 and $e^{+}e^{-}\rightarrow b\bar{b}\Pi^{0}_{t}$. These cross sections
 are
 about two orders or even three orders of magnitude larger than that of
 processes $e^{+}e^{-}\rightarrow f'\bar{f'}\Pi^{0}_{t}$.  The study in this paper is a useful addition to the
 previous studies. Some
 similar processes in the context of the SM and  MSSM have also been
 studied quite extensively in the LC\cite{LC}, Tevatron and
 LHC\cite{LHC}. Reference\cite{Leibovich} has investigated the
 production of the neutral scalar with a pair of top quarks at the hadron collider. They find
 that the neutral scalar may be observed at the LHC via the
 process $e^{+}e^{-}\rightarrow t\bar{t}\phi$. The $q\bar{q}H$
 production mode is extremely interesting for physicists because
 this production mode provides a direct way to measure the Yukawa
 couplings of the quarks with scalar particles, on the other hand, we can detect
 these scalar particles at LC, LHC and Tevatron II with high luminosity through these
 processes.

 The calculation of the production cross sections of the processes
 $e^{+}e^{-}\rightarrow f\bar{f}\Pi^{0}_{t}$ is presented in
 Sec II.
 \section{Cross sections of these processes}
 It is noticeable that the TC2 model may have rich top-quark
 phenomenology since it treats the top quark differently from
 other quarks. The
 couplings of top-pions
 to three family fermions are nonuniversal and the top-pions have
 large Yukawa couplings to the third generation. The Yukawa
 interactions of top-pions to $t,b,c$ quarks can be written
 as\cite{TC2,He}
\begin{eqnarray}\nonumber
\frac{m_{t}\tan\beta} {\upsilon_{\omega}}&
&[iK^{tt}_{UR}K^{tt^{*}}_{UL}\overline{t_{L}}t_{R}
\Pi^{0}_{t}+\sqrt{2}K^{tt}_{UR}K^{bb^{*}}_{DL}\overline{b_{L}}t_{R}\Pi^{-}_{t}\\
\nonumber &
&+iK^{tc}_{UR}K^{tt^{*}}_{UL}\overline{t_{L}}c_{R}\Pi^{0}_{t}+\sqrt{2}K^{tc}_{UR}
K^{bb^{*}}_{DL}\overline{b_{L}}c_{R}\Pi^{-}_{t}\\  &
&+i\frac{m^{*}_{b}}{m_{t}}\overline{b_{L}}b_{R}\Pi^{0}_{t}+h.c.]
\end{eqnarray}
where
$\tan\beta=\sqrt{(\frac{\upsilon_{\omega}}{\upsilon_{t}})-1}$,
$\upsilon_{t}\thickapprox 60-100$ GeV is the top-pion decay
constant and $\upsilon_{\omega}=246$ GeV is the electroweak
symmetry-breaking scale. $m_{b}^{*}$ is the  part of b-quark mass
induced by instanton. $K^{tt}_{UL}, K^{bb}_{DL},
K^{tt}_{UR},K^{tc}_{UR}$ are the elements of the rotation matrices
$K_{UL,R}$ and $K_{DL,R}$. The rotation matrices $K_{UL,R},$ and
$K_{DL,R}$ are needed for diagonalizing the up- and down-quark
mass matrices $M_{U}$ and $M_{D}$, i.e.,
$K_{UL}^{+}M_{U}K_{UR}=M^{dia}_{U}$ and
$K_{DL}^{+}M_{D}K_{DR}=M^{dia}_{D}$, from which the
Cabibbo-Kobayashi-Maskawa(CKM) matrix is defined as
$V=K^{+}_{UL}K_{DL}$. The matrix elements are given as
 \begin{eqnarray*}
K^{tt}_{UL}\backsimeq K^{bb}_{DL}\approx 1 \hspace{1.5cm}
  K^{tt}_{UR}=1-\varepsilon
\end{eqnarray*}
Here, we take $\varepsilon$ as a free parameter changing from 0.03
to 0.1.

 With $t\bar{t}\Pi^{0}_{t}$ coupling,
the neutral top-pion $\Pi^{0}_{t}$ can couple to a pair of gauge
bosons through the top quark triangle loops in an isospin
violating way. Calculating the top quark triangle loops, we can
explicitly obtain the couplings of $\Pi^{0}_{t}-\gamma-\gamma$ ,
   $\Pi^{0}_{t}-\gamma-Z$ and  $\Pi^{0}_{t}-Z-Z$
   \begin{eqnarray}\nonumber
  & & \Pi^{0}_{t}-\gamma-\gamma :\\
  & &  iN_{c}\frac{8}{9\pi}
   \frac{tan\beta}{\upsilon_{w}}m_{t}^{2}(1-\varepsilon)\alpha_{e}
   \varepsilon_{\mu\nu\rho\delta}p_{in}^{\rho}p_{out}^{\delta}C_{0}
   \end{eqnarray}
\begin{eqnarray} \nonumber
 & & \Pi^{0}_{t}-\gamma-Z: \\ \nonumber
  & &iN_{c}\frac{\alpha_e}{3\pi c_w s_w}
   \frac{tan\beta}{\upsilon_{w}}m_{t}^{2}(1-\varepsilon)
   \varepsilon_{\mu\nu\rho\delta}
   p_{in}^{\rho}p_{out}^{\delta}\\
  & & (1-\frac{8}{3}s_{w}^{2})C_{0}
\end{eqnarray}
\begin{eqnarray} \nonumber
 & & \Pi^{0}_{t}-Z-Z: \\ \nonumber
  & &iN_{c}\frac{\alpha_e}{8\pi c_{w}^{2} s_{w}^{2}}
   \frac{\tan\beta}{\upsilon_{w}}m_{t}^{2}(1-\varepsilon)
   \varepsilon_{\mu\nu\rho\delta}
   p_{in}^{\rho}p_{out}^{\delta}\\
  & & \{[(1-\frac{8}{3}s_{w}^{2})^{2}-1]C_{0}-2C_{11}\}
 \end{eqnarray}
 where $N_{c}$ is the color index with $N_{c}=3$, $s_{w}=\sin\theta_{w}$,
 $c_{w}=\cos\theta_{w}$ ($\theta_{w}$ is the Weinberg angle),
   $C_{0}=C_{0}(-p_{in},p_{out},m_{t},m_{t},m_{t})$ and $C_{11}=
   C_{11}(-p_{in},p_{out},m_{t},m_{t},m_{t})$
   are  standard three-point scalar integrals with
   $p_{in}$ and $p_{out}$ denoting the momenta of the incoming gauge boson
   and the outcoming top-pion, respectively.

     With the couplings of $\Pi^0_t\gamma\gamma$,
     $\Pi^0_tZ\gamma$,$\Pi^0_tZZ$, the neutral  top-pion can be produced via the
     processes
     $e^{+}e^{-}\rightarrow
   f\bar{f}\Pi^{0}_{t}$. The Feynman diagrams of these processes are shown in
   Fig.1.
 \begin{figure}[h]
 \begin{center}
  \vspace*{-1.1cm}
 \epsfig{file=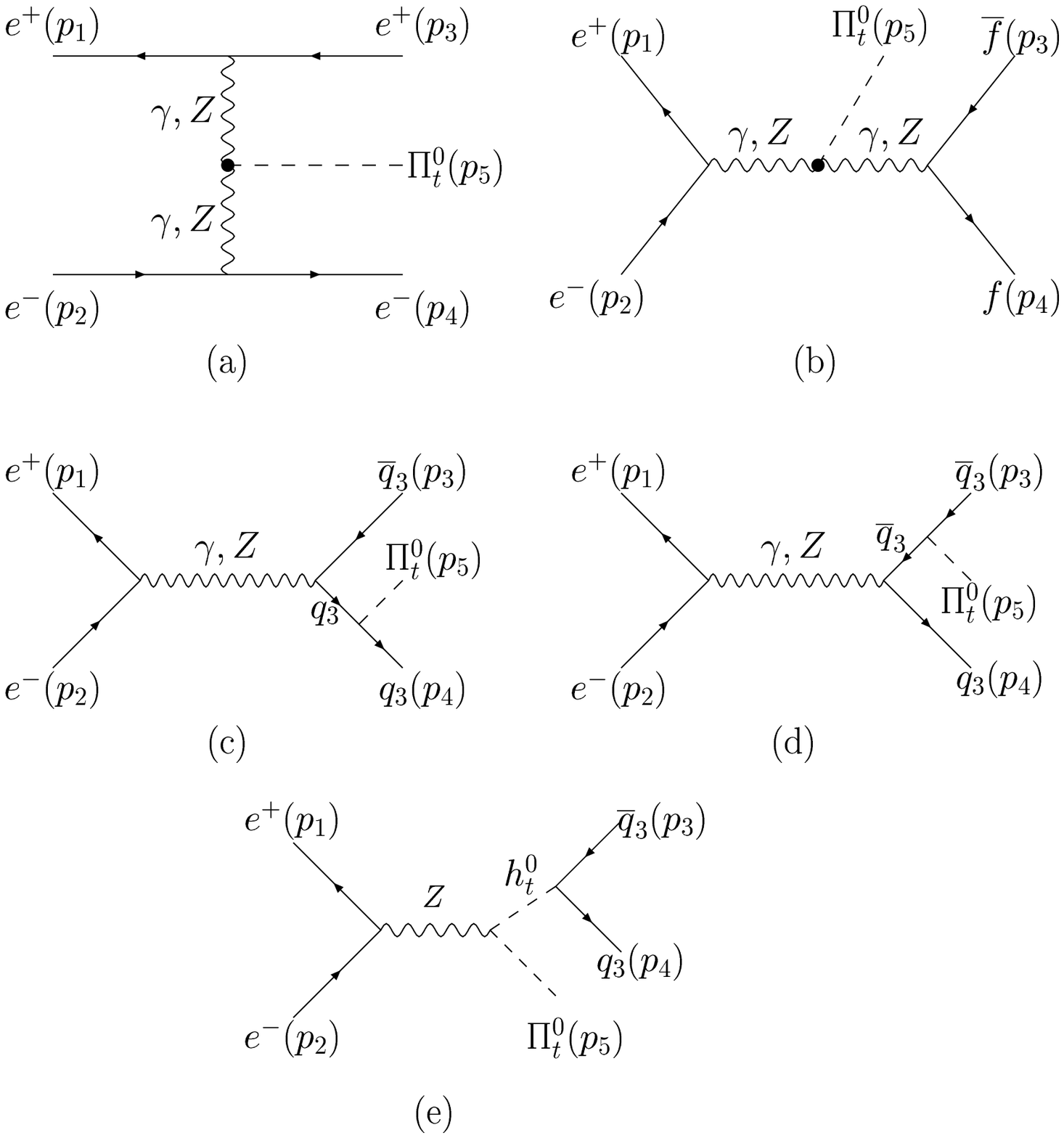,width=260pt,height=360pt} \vspace{-4.0cm}
 \caption{ The Feynman diagrams of the processes
    $e^{+}e^{-}\rightarrow f\bar{f}\Pi^{0}_{t}$. $f$ represents $u,d,c,s,t,b$ quarks and
    $e,\mu,\tau$ leptons. $q_{3}$ represents the third generation quarks $t,b$}
 \label{fig1}
 \end{center}
 \end{figure}
 From the diagrams, we can see that the process  $e^{+}e^{-}\rightarrow e^{+}e^{-}\Pi^{0}_{t}$
 can take place through t-channel and s-channel as shown in Fig.
 1(a),(b). The light quark pairs$(u\bar{u}, d\bar{d}, c\bar{c},s\bar{s})$ or lepton
 pairs $(\mu\bar{\mu}, \tau\bar{\tau})$ can only be produced via Fig.1(b). The heavy quark pair
 production processes $e^{+}e^{-}\rightarrow
 t\bar{t}\Pi^{0}_{t}$ and $e^{+}e^{-}\rightarrow
 b\bar{b}\Pi^{0}_{t}$ are shown in Fig.1(b),(c),(d),(e).
 $h^{0}_{t}$ shown in Fig.1(e) is a CP even particle in TC2 model, the couplings $h^0_tq\bar{q}$ and $Zh^0_t\Pi^0_t$
 are given in Ref.\cite{lugongru}.

  The explicit expressions of the amplitudes for
  different diagrams can be directly written as:
 \begin{eqnarray*}
  M^{a}_{\gamma\gamma\Pi^{0}_{t}}&=&iN_{c}\alpha_{e}^{2}\frac{32}{9}
   \frac{\tan\beta}{\upsilon_{w}}m_{t}^{2}(1-\varepsilon)
   \varepsilon^{\mu\nu\rho\delta}(p_{2}-p_{4})_{\rho}p_{5\delta}
   C_{0}\\
  & & G(p_{2}-p_{4},0)G(p_{2}-p_{4}-p_{5},0)\\
 & & \overline{u}_{e^{-}}(p_{4})\gamma_{\mu}u_{e^{-}}(p_{2})
   \overline{v}_{e^{+}}(p_{1})\gamma_{\nu}v_{e^{+}}(p_{3})\\
 M^{a}_{\gamma Z\Pi^{0}_{t}}&=&-iN_{c}\frac{2^{\frac{5}{4}}}{3}\frac{\alpha_{e}^{2}}
 {c^{2}_{\omega}s^{2}_{\omega}}\frac{\tan\beta}{\upsilon_{w}}m_{t}^{2}(1-\varepsilon)
 \varepsilon^{\mu\nu\rho\delta}(p_{2}-p_{4})_{\rho}p_{5\delta}\\
 & &(1-\frac{8}{3}s^{2}_{\omega})C_{0}G(p_{2}-p_{4},0)G(p_{2}-p_{4}-p_{5},M_{Z})\\
 & &\overline{u}_{e^{-}}(p_{4})\gamma_{\mu}u_{e^{-}}(p_{2})
 \overline{v}_{e^{+}}(p_{1})\gamma_{\nu}(-\frac{1}{2}L+s^{2}_{\omega})
 v_{e^{+}}(p_{3})\\
M^{a}_{Z\gamma\Pi^{0}_{t}}&=&-iN_{c}\frac{2^{\frac{5}{4}}}{3}\frac{\alpha_{e}^{2}}
 {c^{2}_{\omega}s^{2}_{\omega}}\frac{\tan\beta}{\upsilon_{w}}m_{t}^{2}(1-\varepsilon)
 \varepsilon^{\mu\nu\rho\delta}(p_{2}-p_{4})_{\rho}p_{5\delta}\\
 & &(1-\frac{8}{3}s^{2}_{\omega})C_{0}G(p_{2}-p_{4},M_{Z})G(p_{2}-p_{4}-p_{5},0)\\
 & &\overline{u}_{e^{-}}(p_{4})\gamma_{\mu}(-\frac{1}{2}L+s^{2}_{\omega})u_{e^{-}}
 (p_{2})\overline{v}_{e^{+}}(p_{1})\gamma_{\nu}v_{e^{+}}(p_{3})\\
M^{a}_{ZZ\Pi^{0}_{t}}&=&iN_{c}\frac{\alpha_{e}^{2}}{2c^{4}_{\omega}s^{4}_{\omega}}
\frac{\tan\beta}{\upsilon_{w}}m_{t}^{2}(1-\varepsilon)
\varepsilon^{\mu\nu\rho\delta}
(p_{2}-p_{4})_{\rho}p_{5\delta}\\
 &  &\{[(1-\frac{8}{3}s^{2}_{\omega})^{2}-1]C_{0}-2C_{11}\}G(p_{2}-p_{4},M_{Z})\\
 & &G(p_{2}-p_{4}-p_{5},M_{Z})
 \overline{u}_{e^{-}}(p_{4})\gamma_{\mu}(-\frac{1}{2}L+s^{2}_{\omega})\\
 & &u_{e^{-}}(p_{2})
 \overline{v}_{e^{+}}(p_{1})\gamma_{\nu}(-\frac{1}{2}L+s^{2}_{\omega})v_{e^{+}}(p_{3})\\
 M^{b}_{\gamma\gamma\Pi^{0}_{t}}&=&-iQ_{f}N_{c}\alpha_{e}^{2}\frac{32}{9}
 \frac{\tan\beta}{\upsilon_{w}}m_{t}^{2}(1-\varepsilon)
 \varepsilon^{\mu\nu\rho\delta}(p_{1}+p_{2})_{\rho}\\
  & &p_{5\delta} C'_{0}
  G(p_{1}+p_{2},0)G(p_{1}+p_{2}-p_{5},0)\\
 & & \overline{v}_{e^{+}}(p_{1})\gamma_{\mu}u_{e^{-}}(p_{2})
  \overline{u}(p_{4})\gamma_{\nu}v(p_{3})\\
 M^{b}_{\gamma Z\Pi^{0}_{t}}&=&-iN_{c}\frac{2^{\frac{5}{4}}}{3}\frac{\alpha_{e}^{2}}
 {c^{2}_{\omega}s^{2}_{\omega}}\frac{\tan\beta}{\upsilon_{w}}m_{t}^{2}(1-\varepsilon)
 \varepsilon^{\mu\nu\rho\delta}(p_{1}+p_{2})_{\rho}\\
 & &p_{5\delta}(1-\frac{8}{3}s^{2}_{\omega})C'_{0}G(p_{1}+p_{2}-p_{5},M_{Z})\\
  & &G(p_{1}+p_{2},0)
 \overline{v}_{e^{+}}(p_{1})\gamma_{\mu}u_{e^{-}}(p_{2})\\
 & & \overline{u}(p_{4})\gamma_{\nu}(aL+b)v(p_{3})\\
 M^{b}_{Z\gamma\Pi^{0}_{t}}&=&iQ_{f}N_{c}\frac{2^{\frac{5}{4}}}{3}\frac{\alpha_{e}^{2}}
 {c^{2}_{\omega}s^{2}_{\omega}}\frac{\tan\beta}{\upsilon_{w}}m_{t}^{2}(1-\varepsilon)
 \varepsilon^{\mu\nu\rho\delta}(p_{1}+p_{2})_{\rho}\\
 & &p_{5\delta}(1-\frac{8}{3}s^{2}_{\omega})C'_{0}G(p_{1}+p_{2}-p_{5},0)G(p_{1}+p_{2},\\
 & & M_{Z})\overline{v}_{e^{+}}(p_{1})\gamma_{\mu}
 (-\frac{1}{2}L+s^{2}_{\omega})u_{e^{-}}(p_{2})\\
 & &\overline{u}(p_{4})\gamma_{\nu}v(p_{3})\\
 M^{b}_{ZZ\Pi^{0}_{t}}&=&iN_{c}\frac{\alpha_{e}^{2}}{2c^{4}_{\omega}s^{4}_{\omega}}
\frac{\tan\beta}{\upsilon_{w}}m_{t}^{2}(1-\varepsilon)
\varepsilon^{\mu\nu\rho\delta}
(p_{1}+p_{2})_{\rho}p_{5\delta}\\
 &  &\{[(1-\frac{8}{3}s^{2}_{\omega})^{2}-1]C'_{0}-2C'_{11}\}G(p_{1}+p_{2},M_{Z})\\
 & &G(p_{1}+p_{2}-p_{5},M_{Z})
 \overline{v}_{e^{+}}(p_{1})\gamma_{\mu}(-\frac{1}{2}L+s^{2}_{\omega})\\
 & &u_{e^{-}}(p_{2})
 \overline{u}(p_{4})\gamma_{\nu}(aL+b)v(p_{3})\\
 M^{c}_{\gamma}&=&4Q_{f}\pi\alpha_{e}
 \frac{\tan\beta}{\upsilon_{\omega}}m^{*}_{q_{3}}
 G(p_{4}+p_{5},m_{q_{3}})G(p_{1}+p_{2},0)\\
 &  &\overline{u}_{q_{3}}(p_{4})\gamma_{5}(\pslash_{4}+\pslash_{5}+m_{q_3})\gamma^{\mu}
 v_{\bar{q}_{3}}(p_{3})\overline{v}_{e^{+}}(p_{1})\gamma_{\mu}\\
 & &u_{e^{-}}(p_{2})\\
M^{c}_{Z}&=&-4\pi\frac{\alpha_{e}}{s^{2}_{\omega}c^{2}_{\omega}}
 \frac{\tan\beta}{\upsilon_{\omega}}m^{*}_{q_{3}}
 G(p_{4}+p_{5},m_{q_{3}})G(p_{1}+p_{2},\\&  &M_{Z})
 \overline{u}_{q_{3}}(p_{4})\gamma_{5}(\pslash_{4}+\pslash_{5}+m_{q_{3}})\gamma^{\mu}
 (aL+b)\\ & &v_{\bar{q}_{3}}(p_{3})
 \overline{v}_{e^{+}}(p_{1})\gamma_{\mu}(-\frac{1}{2}L+s^{2}_{\omega})
 u_{e^{-}}(p_{2})\\
M^{d}_{\gamma}&=&-4Q_{f}\pi\alpha_{e}
 \frac{\tan\beta}{\upsilon_{\omega}}m^{*}_{q_{3}}
 G(p_{3}+p_{5},m_{q_{3}})\\
 &  &G(p_{1}+p_{2},0)\overline{u}_{q_{3}}(p_{4})\gamma_{5}(\pslash_{3}+\pslash_{5}+m_{q_{3}})
 \\ & &\gamma^{\mu}v_{\bar{q}_{3}}(p_{3})
 \overline{v}_{e^{+}}(p_{1})\gamma_{\mu}u_{e^{-}}(p_{2})\\
M^{d}_{Z}&=&4\pi\frac{\alpha_{e}}{s^{2}_{\omega}c^{2}_{\omega}}
 \frac{\tan\beta}{\upsilon_{\omega}}m^{*}_{q_{3}}
 G(p_{3}+p_{5},m_{q_{3}})G(p_{1}+p_{2},\\
 & &M_{Z})\overline{u}_{q_{3}}(p_{4})\gamma^{\mu}(aL+b)(\pslash_{3}+\pslash_{5}+m_{q_{3}})\\
 & &\gamma_{5}v_{\bar{q}_{3}}(p_{3})\overline{v}_{e^{+}}(p_{1})\gamma_{\mu}
 (-\frac{1}{2}L+s^{2}_{\omega})u_{e^{-}}(p_{2})\\
M^{e}&=&-2\pi\frac{\alpha_{e}}{c_{\omega}s_{\omega}}\frac{\tan\beta}{\upsilon_{\omega}}
 m^{*}_{q_{3}} G(p_{3}+p_{4},m_{h})G(p_{1}+p_{2},\\
 & &M_{Z})\overline{v}_{e^{+}}(p_{1}) (\pslash_{5}-\pslash_{3}-\pslash_{4})
 u_{e^{-}}(p_{2})\\
 & &\overline{u}_{q_{3}}(p_{4})v_{\bar{q}_{3}}(p_{3})
 \end{eqnarray*}
 where
 $L=\frac{1}{2}(1-\gamma_{5})$, $G(p,m)=\frac{1}{p^{2}-m^{2}}$
 denotes the propagator of the particle and
 $C_{0}=C_{0}(-p_{2}+p_{4},p_{5},m_{t},m_{t},m_{t})$,
$C'_{0}=C_{0}(-p_{1}-p_{2},p_{5},m_{t},m_{t},m_{t})$. $m_{q_{3}}$
represents the masses of the third generation quarks and
$m_{q_{3}}^{*}$ denotes $m_{t}^{*}$ and $ m_{b}^{*}$,
$m_{t}^{*}=(1-\varepsilon)m_{t}$  is induced by topcolor
interaction and $m_{b}^{*}$ is the b quark mass produced by
instanton. The values of
$Q_{f},a,b$ are taken as following: \\
\vspace{0.1in}
\begin{center}
\doublerulesep 0.8pt
 \tabcolsep 0.12in
 \begin{tabular}{||c|c|c|c||}\hline \hline
 $ f $ &  $Q_{f}$ & $ a $ & $ b$ \\ \hline
 up-quarks$(u,c,t)$& $\frac{2}{3}$ & $\frac{1}{2}$ &
 $-\frac{2}{3}s^{2}_{\omega}$\\ \hline
 down-quarks$(d,s,b)$ & $-\frac{1}{3}$ & $-\frac{1}{2}$ &
 $\frac{1}{3}s^{2}_{\omega}$\\ \hline
 leptons$(e,\mu,\tau)$ & $-1$ & $-\frac{1}{2}$ & $s^{2}_{\omega}$\\
 \hline \hline
\end{tabular}
\end{center}

The production amplitudes for different processes are:
 \begin{eqnarray*}
M_{e^{+}e^{-}\Pi^{0}_{t}} & = &M^{a}+M^{b} \hspace{2.2cm}
(e^{+}e^{-}\rightarrow e^{+}e^{-}\Pi^{0}_{t})\\
M_{q_{3}\overline{q_{3}}\Pi^{0}_{t}}&=&M^{b}+M^{c}+M^{d}+M^{e}\hspace{0.3cm}
(e^{+}e^{-}\rightarrow q_{3}\overline{q_{3}}\Pi^{0}_{t})\\
M_{f'\overline{f'}\Pi^{0}_{t}}&=&M^{b}(f'=u,d,c,s,\mu , \tau)
\hspace{0.2cm} (e^{+}e^{-}\rightarrow f'\overline{f'}\Pi^{0}_{t})
 \end{eqnarray*}

 With the above production amplitudes, we can obtain the production
cross sections directly.

\section{Numerical results and conclusions}
In our calculation, we take $m_{e}=0$, $m_{\mu}=0.105$ GeV,
 $m_{\tau}=1.784$ GeV, $m_{u}=0.005$ GeV, $m_{d}=0.009$ GeV,
$m_{c}=1.4$ GeV, $m_{s}=0.15$ GeV, $m_{t}=174$ GeV, $m_{b}=4.9$
GeV, $M_{Z}=91.187$ GeV, $v_{t}=60$ GeV and $s^{2}_{\omega}=0.23$.
 The electromagnetic fine-structure constant $\alpha_{e}$ at a
certain energy scale is calculated from the simple QED one-loop
evolution with the boundary value $\alpha_{e}=\frac{1}{137.04}$.
 There are three free parameters $\varepsilon,M_{\Pi},s $ in the cross
 sections. To see the influence of these parameters on the cross sections,
we take the mass of the top-pion $M_{\Pi}$ to vary in the range of
150 GeV$\leq M_{\Pi}\leq$ 450 GeV and $\varepsilon=0.03,0.06,0.1$,
respectively. Considering the center-of-mass energy $\sqrt{s}$ in
the planned $e^{+}e^{-}$ linear colliders(for example,TESLA), we
take $\sqrt{s}=800$ GeV and 1600 GeV, respectively.

For the process $e^{+}e^{-}\rightarrow e^{+}e^{-}\Pi^{0}_{t}$,
 we can see that there exists a t-channel resonance effect induced by the photon propagator in fig.1(a)
 and an s-channel resonance effect induced by the Z boson propagator in fig.1(b). So we should take into account the
effect of the width of the $Z$ boson in the calculations, i.e. we
should take the complex mass term $M_{Z}^{2}-iM_{Z}\Gamma_{Z}$
instead of the simple $Z$ boson mass term $M_{Z}^{2}$ in the $Z$
boson propagator. Here, we take $\Gamma_{Z}=2.49$ GeV. All the
resonance effects will enhance the cross section significantly. We
find that the contribution to the cross section of
$e^{+}e^{-}\rightarrow e^{+}e^{-}\Pi^{0}_{t}$  arises mainly from
Fig.1(a). The final numerical results of the cross section are
summarized in Figs 2. The figure is the plots of the cross section
as a function of $M_{\Pi}$. One can see that there is a peak in
the plot when $M_{\Pi}$ is close to 350 GeV, which arises from the
top quark triangle loop. The largest cross sections are 1.78 fb
and  4.15fb, when we take $\varepsilon=0.03$  and $\sqrt{s}=800$
and 1600 GeV, respectively. The cross section increases with
$\sqrt{s}$. With a luminosity of $100fb^{-1}/$yr, there are
several tens or even hundreds of $\Pi^{0}_{t}$ events to be
produced via the process $e^{+}e^{-}\rightarrow
e^{+}e^{-}\Pi^{0}_{t}$. For $M_{\Pi}\leq$ 350 GeV, the dominate
decay channel of $\Pi^{0}_{t}$ is $\Pi^{0}_{t}\rightarrow
t\overline{c}$.  As  has been investigated in Ref. \cite{yue}, the
\noindent branching ratio Br$(\Pi^{0}_{t}\rightarrow
t\overline{c})$ can reach about 60\%. Because there is no tree
level flavor-changing neutral current in the SM, the background of
$e^{+}e^{-}\rightarrow e^{+}e^{-}\Pi^{0}_{t}\rightarrow
e^{+}e^{-}t\bar{c}$ is very clean. Therefore,
$e^{+}e^{-}\rightarrow e^{+}e^{-}\Pi^{0}_{t}\rightarrow
e^{+}e^{-}t\bar{c}$ is an ideal channel to detect the neutral
top-pion with small top-pion mass. For $ M_{\Pi}\geq$ 350 GeV,
$\Pi^{0}_{t}\rightarrow t\overline{t}$ is permitted and the total
width of $\Pi^{0}_{t}$ increases significantly. The branching
ratio of the decay channel $\Pi^{0}_{t}\rightarrow t\overline{t}$
is close to 100\%, all other decay modes may be ignored. In this
case, we should detect $\Pi^{0}_{t}$ through the $t\bar{t}$
channel.

\vspace*{-0.6cm}
\begin{figure}[h]
\begin{center}
\epsfig{file=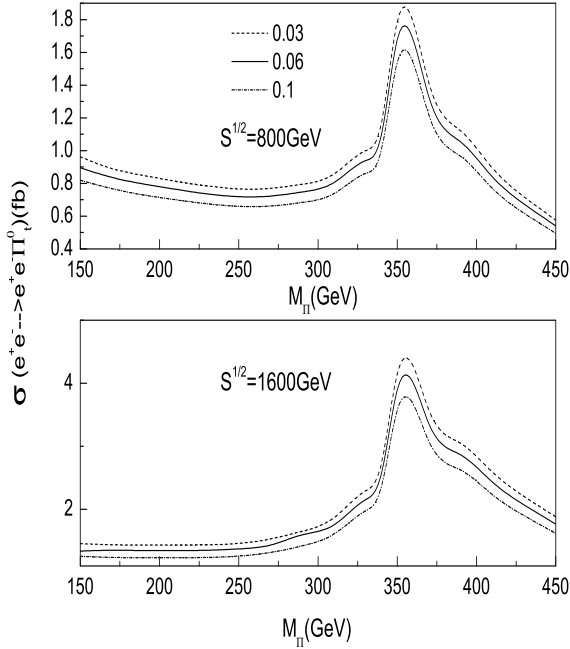,width=240pt,height=280pt} \caption{The cross
section of $ e^{+}e^{-}\rightarrow  e^{+}e^{-}\Pi^{0}_{t}$ versus
top-pion mass $M_{\Pi}$(150-450 GeV) for $\sqrt{s}=800$,1600 GeV
and $\varepsilon=0.03$(dash line),$\varepsilon=0.06$ (solid line
),$\varepsilon=0.1$(dash-dot line) repectively.} \label{fig2}
\end{center}
\end{figure}

\vspace*{-1.2cm}
\begin{figure}[h]
\begin{center}
\epsfig{file=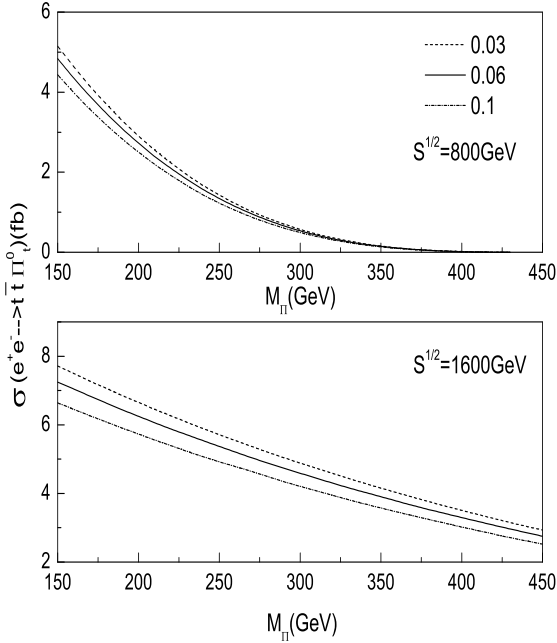,width=240pt,height=280pt} \caption{The same
plots as Fig.2 for the process of  $e^{+}e^{-}\rightarrow
t\overline{t}\Pi^{0}_{t}$. } \label{fig3}
\end{center}
\end{figure}

The numerical results of the process $e^{+}e^{-}\rightarrow
t\overline{t}\Pi^{0}_{t}$ are shown in Fig.3. The contributions
arise from Fig.1(b)(c)(d)(e). There is no resonance effect for
this process, but the large $\Pi_t^0t\bar{t}$ coupling in
Fig.1(c)(d) can enhance the cross section significantly. We can
see that the cross section increases with $M_{\Pi}$ decreasing and
$\sqrt{s}$ increasing. In the all parameter space we consider, the
cross section is larger than 2.5fb for $\sqrt{s}=1600$ GeV.

The same plots as Fig.2 for the process $e^{+}e^{-}\rightarrow
b\bar{b}\Pi^{0}_{t}$ are shown in  Fig.4. The contributions to the
cross section  come mainly from Fig.1(e) due to the resonance
effect of the $h^{0}_{t}$ propagator. In this case, the effect of
the decay width of $h^0_t$ should be considered. The possible
decay modes of $h^{0}_{t}$ are $b\bar{b},t\bar{c},gg,W^+W^-,ZZ,$
$t\bar{t}$(if $M_{h^0_t}\geq 2m_{t}$). The results show that
although the coupling of $\Pi_t^0b\bar{b}$ is much smaller than
the coupling of $\Pi_t^0t\bar{t}$, the resonance effect can
enhance the cross section to the level of a few fb.  When
$\sqrt{s}=1600$ GeV, the cross section will be much smaller than
that for $\sqrt{s}=800$ GeV and it hardly varies with $M_{\Pi}$.
So we only draw up the plots for the case of $\sqrt{s}=800$ GeV.
The plots show that the cross section is not sensitive to
$M_{\Pi}$. The neutral top-pion can be more easily detected via
the process $e^{+}e^{-}\rightarrow b\bar{b}\Pi^{0}_{t}$ than via
the process $e^{+}e^{-}\rightarrow t\bar{t}\Pi^{0}_{t}$ with
b-tagging. Therefore, the process $e^{+}e^{-}\rightarrow
b\bar{b}\Pi^{0}_{t}$ provides us with another useful way to search
for a neutral top-pion.

For the other processes $e^{+}e^{-}\rightarrow
f'\bar{f'}\Pi^{0}_{t}(f'=u,d,c,s,\mu,\tau)$, the largest cross
section could only be up to 0.01 fb. Therefore, we could hardly
detect the neutral top-pion through these processes. So we do not
discuss these processes in detail.

\vspace*{-0.8cm}
\begin{figure}[h]
\begin{center}
\epsfig{file=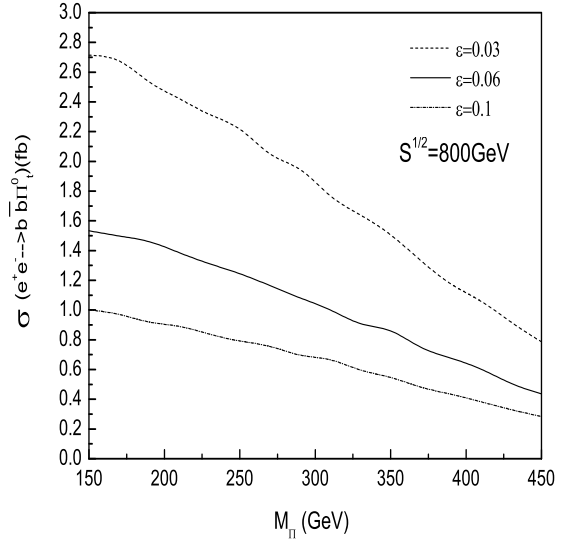,width=240pt,height=240pt} \caption{The same
plots as Fig.2 for the process of $e^{+}e^{-}\rightarrow
b\overline{b}\Pi^{0}_{t}$ when $\sqrt{s}=800$ GeV} \label{fig4}
\end{center}
\end{figure}

 In conclusion, we have studied some  neutral top-pion
 production processes
 $e^{+}e^{-}\rightarrow f\bar{f}\Pi^{0}_{t}(f=u,d,c,s,t,b,e,\mu,\tau)$ at the future
 $e^{+}e^{-}$ colliders in the framework of the TC2 model. We find that there are the following features
 for these processes:
(i)The cross sections  of the processes $e^{+}e^{-}\rightarrow
f'\bar{f'}\Pi^{0}_{t}(f'=u,d,c,s,\mu,\tau)$ are too small to
detect a neutral top-pion. (ii)Due to the effect of the top quark
triangle loops, there exists a narrow peak in the cross section
plots of the process $e^{+}e^{-}\rightarrow
e^{+}e^{-}\Pi^{0}_{t}$. Because of the resonance effect, the cross
section of $e^{+}e^{-}\rightarrow e^{+}e^{-}\Pi^{0}_{t}$ could be
up to a few fb and a few tens or even hundreds of $\Pi^{0}_{t}$
events can be produced, which causes  the neutral top-pion to
become experimentally detectable through the flavor-changing
$t\bar{c}$ channel with the clean background. (iii)The cross
sections of both $e^{+}e^{-}\rightarrow b\overline{b}\Pi^{0}_{t}$
and $e^{+}e^{-}\rightarrow t\overline{t}\Pi^{0}_{t}$ can reach the
level of a few fb, the strong coupling $t\bar{t}\Pi_t^0$ in the
process $e^{+}e^{-}\rightarrow t\overline{t}\Pi^{0}_{t}$ and the
resonance effect of the $h^0_t$ propagator in the process
$e^{+}e^{-}\rightarrow b\overline{b}\Pi^{0}_{t}$ could enhance the
cross sections significantly. These two processes provide us with
another way to search for a neutral top-pion. Therefore, our
studies
 could provide a direct way to test the TC2
model by detecting top-pion signals.

 \vspace*{0.5cm}
\noindent{\bf Acknowledgments}

This work is supported by the National Natural Science Foundation
of China(10175017 and 10375017), the Excellent Youth Foundation of
Henan Scientific Committee(02120000300), and the Henan Innovation
Project for University Prominent Research Talents(2002KYCX009).

%%%%%%%%%%%%%%%%%%%%%%%%%%%%%%%%%%%%%%%%%%%%%%%%%%%%%%%%%%%%%%%%%%%%%%%
%\begin{figure}[htb]
%\begin{center}
%\epsfig{file=mtpg.eps,width=240pt,height=280pt} \caption{The
%toppion-charm invariant mass distribution for $\sqrt{s}=500$ GeV
 %$\varepsilon=0.1$ and $M_{\Pi}=150$ GeV.}
%\label{fig5}
%\end{center}
%\end{figure}
%%%%%%%%%%%%%%%%%%%%%%%%%%%%%%%%%%%%%%%%%%%%%%%%%%%%%%%%%%%%%%%%%%%%%%%%%%%%%%%%%%%%%%%%%%%%%%
\vspace*{-.2cm}
%%%%%%%%%%%%%%%%%%%%%%%%%%%%%%%%%%%%%%%%%%%%%%%%%%%%%%%%%%%%%%%%%%%%%%%%%%%%%%%%%%%%%
\null

\end{document}